\documentstyle[preprint,aps,twoside,epsf]{revtex}

\begin{document}

\draft

\title{Nucleation of Superconducting pairing states at mesoscopic scales
at zero temperature}

\author{F.Zhou$^{a}$, Cristiano Biagini$^{b}$}
 
\address{$^{a}$Physics Department, Princeton University,
Princeton, NJ 08544}
\address{$^{b}$INFM, Unita'di Napoli, Mostra d'Oltremare, Pad.19, 80125,
Napoli, Italia}

\maketitle

\begin{abstract}
We find that spin polarized disordered Fermi
liquids are unstable to the nucleation
of superconducting pairing states at mesoscopic
scales even in magnetic fields substantially
higher than the critical one. 
We study the probability of finding 
superconducting pairing states at mesoscopic scales in 
this limit.
We find that the distribution function  
depends only on the film conductance.
The typical length scale at which pairing takes
place is universal, and decreases when the magnetic field
is increased. 
The number density of these states determines the strength
of the random exchange interactions 
between mesoscopic pairing states.

\end{abstract}


\newpage
\narrowtext

The stability of a superconducting state
with an order parameter $\Delta$ can be characterized
in term of the generalized curvature in the 
following way,

\begin{eqnarray}
{\cal O}({\bf r}, {\bf r}')=\frac{\delta^2 E(\{\Delta({\bf r})\})}
{\delta \Delta^{*}({\bf r})\delta \Delta({\bf r}')}.~
\end{eqnarray}
The curvature determines the stability to the 
spatial variation of the modulas of the order parameter
and the 
stability to the creation of supercurrents.
Here $E(\{\Delta({\bf r})\})$ is the energy
of a configuration $\{ \Delta({\bf r})\}$.
The curvature evaluated at the ground state should 
be positive defined, i.e. $ det~{\cal O}({\bf r}, {\bf r}') > 0 $. 
For a dirty superconductor where
$\sqrt{D/\Delta} \gg l \gg p_F^{-1}$, 
the curvature at $\Delta({\bf r})=\Delta$
has mesoscopic fluctuations, like other physical 
quantities$^{[1]}$. 
Here $D$ is the diffusion constant and 
$l$ is the mean free path, $p_F$ is the Fermi momentum.
However, in the absence of an external magnetic field,
the mesoscopic fluctuations are small.
Thus the curvature is 
almost positive defined, and the conventional homogenous superconducting state
is stable.

When the magnetic field is applied parallel
to the disordered thin superconducting film,  the  
suppression of superconductivity is mainly due to Zeeman 
splitting of electron spin energy levels$^{[2,3]}$. 
In the strong spin-orbital scattering limit  
$\tau_{so}\Delta_0 \ll 1$, close to the critical field $H_c^0$,  
the average spin polarization energy $E_{p}$ 
is nearly equal to the average condensation energy $E_c$. 
Mesoscopic fluctuations of spin polarization energy 
due to mesoscopic fluctuations of spin sucseptibility 
become comparable or larger than
the energy difference $E_c - E_p$. 
Therefore, the amplitude of the mesoscopic 
fluctuations of the curvature evaluated at
$\Delta=\Delta(H)$ could be comparable with the average. 
Here $\Delta(H)$ is the order parameter at given $H$, $\Delta_0$ is the order
parameter for $H=0$ and $1/\tau_{so}$ is the spin-orbit scattering rate.
At $H_{c}^{0} -H/H^{0}_{c}\sim g^{-2}$, where $g=e^2\nu_0Dd/(2\pi^2\hbar)$
is the dimensionless film conductance, $\nu_0$ the density of states and $d$
the film thickness, the curvature, averaged over the area of size 
$\xi(H)=\sqrt{D\Delta_0/\Delta^2(H)}$, 
has a random sign, and the system is unstable with
respect to the creation of normal regions or
spontaneous creation of supercurrents$^{[4]}$.

One of the consequences of the mechanism discussed above
is the {\em instability of the spin polarized disordered Fermi liquid}
well above the critical magnetic field. 
This is because though the average curvature of the normal metal state 
(${\cal O}({\bf r}, {\bf r}')$ evaluated at $\Delta=0$)is positive defined,
i.e. $E_p > E_c$, its mesoscopic fluctuations have 
random signs because of the mesoscopic
fluctuations of the spin polarization energy. 
In the regions where the spin polarization energy cost to form
superconducting pairing state is much lower than the average
energy cost,
the fluctuations of the curvature
are of large negative value comparable to its positive average such that
the normal metal with $\Delta=0$ becomes unstable.  As a result,  
above the critical field $H_c^0$, the superconducting pairing 
correlations are established at mesoscopic scales 
in the different regions in the normal metal and  
couple with each other via exchange interactions of random signs.

In this paper, 
we study the probability to find regions  
where the superconducting pairing states are formed at 
mesoscopic scales at $H > H_c^0$.  
At high magnetic fields
in the strong spin-orbit scattering limit, the statistics of these pairing 
states can be studied with the help of the generalized 
Landau-Ginsburg equation, 
\begin{equation} 
\left[\xi_{0}^{2}\left(\nabla-i\frac{2e}{c}{\bf A}\right)^2 + \frac{H^0_c
-H}{H^0_c}\right]\Delta({\bf r})+ 
\int \delta {\cal O}({\bf r}, {\bf r}',H)\Delta({\bf r}')d{\bf r}'
=\frac{\Delta^3({\bf r})}{2\Delta^2_0},
\end{equation}
where $\xi_0=\sqrt{D/\Delta_0}$.
The statistical property 
of the random potential $\delta {\cal O}({\bf r}, {\bf r})$ 
is determined by its second moment in the Gaussian approximation$^{[4]}$

\begin{equation}
\left\langle\delta {\cal O}({\bf r}_1, {\bf r}'_1)
\delta {\cal O}({\bf r}_2, {\bf r}'_2)\right\rangle
\propto \frac{1}{g^2}
\left[\xi_0^2\delta({\bf r}_1-{\bf r}_2)\delta({\bf r}'_1 -{\bf r}'_2) 
\delta({\bf r}_1-{\bf r}'_1)
+\delta({\bf r}_1-{\bf r}'_2)\delta({\bf r}'_1 -{\bf r}_2) 
\frac{\xi^4_0}{|{\bf r}_1-{\bf r}'_1|^4}\right]
\end{equation}
where
$\left\langle...\right\rangle$ denotes the average over
the impurities realizations.
The curvature evaluated at the normal metal state
where $\Delta({\bf r})=0$ has a simple form:
\begin{eqnarray}
{\cal O}({\bf r}, {\bf r}')=-\left[\xi_0^2\left(\nabla 
- i\frac{2e}{c}{\bf A}\right)^2
+\frac{H_c^0 -H}{H_c^0}\right]\delta({\bf r}-{\bf r}')
+\delta {\cal O}({\bf r}, {\bf r}')
\end{eqnarray}

Eq.2 is a {\em nonlinear} equation in terms of $\Delta({\bf r})$,
with a {\em nonlocal}  
$\delta {\cal O}({\bf r}, {\bf r}')$ potential
originating from the oscillations of the wave functions
of cooper pairs. 
Generally speaking, it is
qualitatively different from the 
Schroedinger equation of an electron in the presence 
of random impurity potentials$^{[5-8]}$.
These complications arise naturally in the study of the 
interplay between the mesoscopic effects and the superconductivity 
and are the generic features of {\em strongly correlated} mesoscopic
systems. In fact, this nonlocal structure of the potential
in Eq.2 leads to the superconducting glass state. 

However, at $H-H_c^0 \gg H_c^0/g^2$, the 
optimal configurations which determine the
macroscopic properties of the sample
turn out to be the superconducting droplets embedded inside
the disordered Fermi liquid, with the phases of each 
droplet coupled via random exchange interaction. 
Such a configuration can be characterized by three parameters:
A). the typical size of the droplet, $L_f$;
B). the typical distance between the droplets, $L_d \gg L_f$;
C). the typical value of the order parameter inside each droplet.
In the following, we will discuss the 
statistical of the mesoscopic pairing states in this regime. 
In this limit, in the leading
order of $(L_f/L_d)^2$ the statistical property of the formation of
superconducting pairing states at mesoscopic scales 
is similar to that of  
the impurity band tails$^{[5-8]}$. 

Eq.2 has a nonzero solution only in the regions where the fluctuation of
the curvature is of order of average, i.e. ${H-H_c^0}/H_c^0$
and the corresponding curvature at $\Delta=0$ is negative.
The nonlinear term in Eq.2 determines the amplitude of $\Delta$ 
and its effect will be discussed only when the typical amplitude of
$\Delta$ is concerned. 
The most probable configurations are those with
superconducting droplets embedded inside the normal metal, i.e. 
$\Delta({\bf r})=\sum_{\alpha}\Delta_\alpha \eta_\alpha({\bf r}),
\int d{\bf r} \eta_\alpha({\bf r})\eta_\beta({\bf r})\propto 
\delta_{\alpha\beta}$.
Note $\eta({\bf r})$ introduced in this way is dimensionless.
For such a configuration to have lower energy than the normal 
state,  $\int d{\bf r} d{\bf r}' \Delta({\bf r}){\cal O}({\bf r}, 
{\bf r}')\Delta({\bf r}') <0$. 
The total energy of such a configuration consists of terms
corresponding to the coupling between
different droplets. 
The coupling between the droplets decays as distance
increases.
When the size of the droplets is much smaller than the distance 
between them,
the typical magnitude of the coupling
between different droplets is much smaller than that of
the coupling within one droplet, provided $L_d \gg L_f$.   
We are going to neglect such terms in 
the estimate of the probability of the droplets
in the leading order of $o(L^2_f/L^2_d)$. 
Thus, to have $l$ droplets in the normal metal,   
$l$ independent inequalities have to be satisfied

\begin{equation}
\Delta^2_\alpha \left[
\int d{\bf  r}\eta_\alpha({\bf r})\left(\xi_0^2\nabla^2 +\frac{H-H^0_c}{H^0_c}
\right)
\eta_\alpha({\bf r}) + 
\int d{\bf r} d{\bf r}'  \eta_\alpha({\bf r})
\delta {\cal O}({\bf r}, {\bf r}') \eta_\alpha({\bf r}')\right] < 0 
\end{equation}
Furthermore, we can write down the probability to have
superconducting pairing states at $H \gg H_c^0$ in term of the sum
of probability to have certain number of droplets 
${\cal P}(\{\eta(x)\})=\sum_{l} P_l(\{\eta_\alpha\}|\alpha=1,...,l)$.
To simplify the notation, we introduce
$O_{LG}=\xi_0^2 \nabla^2 +{H- H_c^0}/{H_c^0},
K_M=\delta {\cal O}({\bf r}, {\bf r}')$. 
Taking into account
$D\eta({\bf r})=\Pi_\alpha D\eta_\alpha$,
we have 

\begin{eqnarray}
P_l(\{\eta_\alpha\}|\alpha=1,...,l)
=\int P(\{K_M\}) \Pi_\alpha 
N^l \int \theta(-L_\alpha+F_\alpha) 
D\eta_\alpha DK_M
\label{probdrop}
\end{eqnarray}
where
$L_\alpha(\{\eta({\bf r})\})=
\int d{\bf r} \eta_\alpha({\bf r}) {O_{LG}} \eta_\alpha({\bf r})$, $  
F_\alpha(\{\eta({\bf r})\})=\int d{\bf r} d{\bf r}' 
\eta({\bf r})_\alpha K_M({\bf r}, {\bf r}') \eta_\alpha({\bf r}')$,
and $N$ is a normalization constant.
We use the following equality to transform the step function 
into integrals: 

\begin{equation}
\theta(-L_\alpha+F_\alpha)=
\int_{-\infty}^{0} dg_\alpha 
\int_{-\infty}^{+\infty}dh_\alpha 
\exp\left[ih_\alpha \left(L_\alpha-F_\alpha - g_\alpha\right)\right]. 
\end{equation}
Eq. 6 is reduced to
$P_l(\{\eta_\alpha\}|\alpha=1,...,l)
=\Pi_{\alpha}\rho_\alpha$. 
In the Gaussian approximation, when the statistics of 
$\delta {\cal O}({\bf r}, {\bf r}')$ is completely determined 
by the second moment of the correlation function, 
$\rho_\alpha$ can be simplified in closed form as 

\begin{equation}
\rho_\alpha
= N \int {\rm erfc}\left[   
\frac{\int d{\bf r} \eta_\alpha({\bf r}) O_{LG}\eta_\alpha({\bf r})}  
{\sqrt{\int d{\bf r}_1 d{\bf r}'_1  \int d{\bf r}_2 d{\bf r}'_2  
{\cal C}({\bf r}_1, {\bf r}'_1; {\bf r}_2, {\bf r}'_2) 
\eta_\alpha({\bf r}_1) \eta_\alpha({\bf r}'_1) 
\eta_\alpha({\bf r}_2) \eta_\alpha({\bf r}'_2) }}\right]
D\eta_\alpha
\end{equation}
where ${\cal C}({\bf r}_1, {\bf r}'_1; {\bf r}_2, {\bf r}'_2)=
\left\langle\delta{\cal O}\delta{\cal O}\right\rangle$ 
as given in Eq.3.
${\rm erfc}({a}/{b})=\int_a\exp(-x^2/2b^2)/\sqrt{2\pi b^2} dx$.
One can evaluate the functional integral $D\eta_\alpha({\bf r})$ in the saddle
point approximation as long as $H-H_c^0 \gg H_c^0/g^2$. 
The saddle point equation of Eq.8 can be obtained by
minimizing the argument of the error function. 
The solution of the saddle
point equation $\eta_s({\bf r})$ determines the shape of the optimal droplets. 
To carry out the functional integral of $\eta_\alpha({\bf r})$, one can expand
$\eta({\bf r})$ around the saddle point,
$\eta({\bf r})=\eta_{s}({\bf r}) + \delta \eta({\bf r}),
\delta \eta({\bf r})=\sum_{n} a_n \eta_n({\bf r})$,
where $\eta_n({\bf r})$ are the eigenstates of the operator 
$\Gamma({\bf r}, {\bf r}')$ generated via second functional
derivative of the argument in the error function with respect to
$\eta({\bf r})$ at $\eta({\bf r})=\eta_s({\bf r})$. Our final result 
depends slightly on the detail structure of $\Gamma({\bf r}, {\bf r}')$ 
and we do not
give an explicit form here.
Performing the gaussian integral of $\delta \eta({\bf r})$
around the saddle point, taking into account the normalization condition,
we obtain,

\begin{equation}
\rho_\alpha={\rm erfc}(\frac{L_{s}}{\sqrt{2\sigma_{s}}})
\frac{det' \Gamma({\bf r}, {\bf r}')}
{det\left\langle{\cal O}({\bf r}, {\bf r}')
\right\rangle}
\int [da_0]
\end{equation} 
where
$L_{s}/{\sqrt{2 \sigma_{s}}}$ is the argument of
error function in Eq.8 evaluated at $\eta({\bf r})=\eta_s({\bf r})$.
$'$ indicates the exclusion of the zero eigenvalue. 
The last integral in Eq.9 corresponds to the contribution from
the zero eigenvalue state, originating from
the translation invariance of the saddle point equation, 
with $2-fold$ degeneracy   
$\eta_{0i}({\bf r})=L_0 \partial_{{0i}} \eta_s({\bf r}-{\bf r}_0)
/\sqrt{\int \eta_s^2 d{\bf r}}, 
i=x,y$$^{[6,7]}$. 
Here $L_0$ is the characteristic length of the droplets determined via
the normalization condition
$1={\int d{\bf r} L_0^2 (\nabla \eta_s)^2}
/{2\int \eta^2_s d{\bf r}}$.
Thus,
\begin{equation}
\int [da_0]=\frac{1}{L_0^2}\int_{v_a} dx_\alpha dy_\alpha. 
\end{equation}
The spatial integral is performed only in the region $v_\alpha$ where
no other droplets are present.  Using the following rescaling 

\begin{eqnarray}
{\bf r}={y}{L_f},~
\nabla=\nabla_y L_f^{-1},~
\eta_s({{\bf r}})=\eta_s(\frac{y}{L_f}),~
{\cal C}({\bf r}, {\bf r}';{\bf r}_1, {\bf r}'_1)=\frac{1}{g^2} 
\frac{\xi_0^2}{L_f^6}  
\tilde{\cal C}(y, y'; y_1, y'_1),~
L_f=\xi_0(\frac{H^0_c}{H -H^0_c})^{1/2},
\end{eqnarray}
we can express $L_{s}$, $\sigma_{s}$
in term of dimensionless $\eta_s(y)$ 
 
\begin{eqnarray}
&&{L_{s}}
=B\frac{H-H^0_c}{H^0_c}L_f^2,~
\sigma_{s}=A^2\frac{\xi_0^2 L_f^2}{g^2}
\nonumber \\
&&B=\int dy {\eta_s}(y)(\nabla^{2}+ 1){\eta_s}(y),
A^2={\int dy_1 dy'_1 dy_2 dy'_2
\tilde{\cal C}(y_1, y'_1; y_2, y'_2){\eta_s}(y_1) {\eta_s}(y'_1)
{\eta_s}(y'_2){\eta_s}(y_2)}
\end{eqnarray}
where $B, A^2$ are the dimensionless quantities
of order of unity depending on the details of ${\eta_s(y)}$. 
$\eta_s$ satisfies the dimensionless saddle point equation 
\begin{eqnarray}
(\nabla^{2}_y+1){\eta_s}(y)
+ \int dy_1dy'_1 dy'  
\tilde{\cal C}(y, y'; y_1, y'_1)
\eta_s(y_1)\eta_s(y'_1)\eta_s(y')=0 
\end{eqnarray}
and at $y=\infty$, $\eta_s(y)=0$. 
We also estimate 
that $det'~ \Gamma({\bf r}, {\bf r}')/det\left\langle{\cal O}({\bf r}, {\bf r}')
\right\rangle
\sim 1$.

Collecting all the results, we have 
\begin{equation}
P_l(\{\eta_\alpha\}|\alpha=1,...,l)
\propto \frac{V^l}{l!}  
\left\{\frac{1}{ L_f^2}
{\rm erfc}\left[\frac{{B g}}{A}\left(\frac{H_c -H}{H_c}\right)^{1/2}\right]
\right\}^{l}
\end{equation}
where $V^l/l!$ is from the spatial integral in Eq.10, excluding the 
overlap between different droplets. We take into account $L_0\sim L_f$.
It is easy to confirm that the average number density of the droplets
is

\begin{equation}
\rho=\frac{1}{V}\frac{\sum_l P_l l}{P_l}\propto \frac{1}{L_f^2}
{\rm erfc}\left[\frac{B g}{A} \left(\frac{H -H^0_c}{H^0_c}\right)^{1/2}\right].
\end{equation}

Let us turn to the problem of the typical amplitude of
$\Delta_\alpha$ of a droplet. The amplitude of 
$\Delta_\alpha$ is determined by the
nonlinear term of Eq.2 and the probability to have a superconducting droplet 
with $\Delta_\alpha=\Delta$ is

\begin{equation}
{\cal P}(\Delta)=N \frac{{2\Delta N_\alpha}}{\Delta^2_0}
\int P({K_M})\delta\left[N_\alpha \left(\frac{\Delta}{\Delta_0}\right)^2
+L_\alpha -F_\alpha\right]
dK_M d\eta_\alpha
\end{equation} 
where $N_\alpha$ is given as
$N_\alpha=\int 
\eta^4_\alpha({\bf r}) d{\bf r}$. $L_\alpha, F_\alpha$ are given 
after Eq.6.
The prefactor in front of the integral is from the Jacobian 
under the transformation $\delta(x) \rightarrow 2Ax \delta(Ax^2)$. 
Transforming $\delta$-function into a integral and carrying out the
gaussian integral of $K_M$ , using 
the procedure similar to that used in obtaining Eq.8, 
we can expressed Eq.16 
in terms of $\eta_\alpha({\bf r})$. The functional integral
can be performed in the saddle point approximation.  
This saddle point equation turns out to be the same as that of Eq.8 except
there is an additional nonlinear term proportional to 
$N_\alpha(\Delta^2/\Delta_0^2)$.
However, as shown below,
the typical $\Delta$ in the optimal droplet is 
much smaller than $\Delta_0\sqrt{H-H_c^0/H_c^0}$
in the limit $H-H_c^0/H_c^0 \gg 1/g^2$. Therefore,
this nonlinear term is much smaller than the linear
term already present in Eq.13
and can be
treated as a perturbation 
as far as the spatial dependence is concerned. 
As a result,
we can use the saddle point solution obtained in Eq.13 to evaluate Eq.16.
By expanding the resultant equation in term of the nonlinear term
$(\Delta/\Delta_0)^2$ and keeping only the $\Delta$ dependent term,
we obtain the {\em conditional}
distribution function of $\Delta$ of a droplet

\begin{equation} 
{\cal P}_c(\Delta)=\frac{2Cg^2 \Delta}{\Delta^2_0}
\exp\left(-C g^2\frac{\Delta^2}{\Delta^2_0}\right).
\end{equation}
This shows that $\Delta \sim \Delta_0/g$ is independent  
of the magnetic field. $C$ is a constant of order of unity.
It is much smaller than $\Delta_0 
\sqrt{H-H_c^0/H_c^0}$, justifying the approximation we made to 
derive Eq.17.

So far we neglect the coupling between different droplets and 
treat the droplets as a 
dilute gas. Though the coupling between droplets 
does not affect the probability of finding
one droplet, it determines the coupling between 
the different droplets
and the global phase rigidity.
The typical distance $L_d$ is order of $L_f 
{\rm erfc}^{-1/2}(Bg/A (H-H_c^0/H_c^0)^{1/2})$ following Eq.15. 
The typical coupling between two droplets is determined by
$\delta {\cal O}$ and is given as 

\begin{equation}
\left|\nu_0 \Delta_\alpha \Delta_\beta 
\int d{\bf r} d{\bf r}' {\cal O}({\bf r},
{\bf r}')\eta_s({\bf r}-{\bf r}_\alpha)
\eta_s({\bf r}' - {\bf r}_\beta)\right| \propto \frac{\Delta_0}{g}
{\rm erfc}\left[\frac{Bg}{A}\left(\frac{H-H_c^0}{H_c^0}\right)^{1/2}\right].
\end{equation}
To obtain this result, we take into account that the size of the droplet
is $L_f$, typical $\Delta_\alpha$ is given by Eq.17, 
and $|{\bf r}_\alpha -{\bf r}_\beta| \sim L_d$. 
The average ${\cal O}$, as shown in Eq.3
is proportional to $\delta({\bf r}-{\bf r}')$.
Thus, the average coupling
is proportional to the overlap integral  
$\int d{\bf r} \eta_s({\bf r}-{\bf r}_\alpha)
\eta_s({\bf r} -{\bf r}_\beta) \propto \exp(-L_d/L_f)$, which is
small in the limit $L_d \gg L_f$.  
This indicates that the distribution function
of the coupling between different pairing states is symmetric with 
respect to zero.

The existence of random Josephson coupling in the 
presence of a parallel magnetic field is a consequence
of the Pauli spin polarization. This phenomena exists 
even without spin orbit scattering. Consider
for example a granular superconductor,
with superconducting grains coupled with each other
via Josephson coupling. 
The sign of the Josephson coupling
is determined by the total phase of the time reversal
pairs. In the pure limit, though the sign of the 
wave function of each electron oscillates with 
a period of Fermi wave length, the total
phase of $(\bf p, -\bf p)$ pair
is zero because of the exact cancellations of the 
phases of each electron inside the pair. 
Therefore there is no sign oscillation
for Josephson couplings.
In the dirty case $\bf p$ is not a good quantum number.
However the sign of the coupling doesn't oscillate as a
function of spatial coordinate because of the time reversal
symmetry. As a result, even 
when the distance between the grains is much larger than
the mean free path, 
the sign of the coupling is positive definite$^{[9]}$. 
This is in contrast to the ferromagnetic ordering
nuclear spins due to RKKY exchange interaction.
In the pure limit, RKKY coupling exhibits
Friedel oscillations with the period of Fermi wave length.
In the presence of impurity scatterings, the phase of Friedel
oscillations of electron wave functions becomes
random. As a result, the nuclear spin system exhibits spin-glass
type ordering instead of ferromagnetic 
ordering when the impurities are present.

In the presence of a parallel magnetic field,
the electrons inside the normal metal become polarized. 
In this case, the electron with spin up
has a different energy as the electron with spin down at the Fermi
surface because of the Pauli spin polarization. As a result,
the phase of the electron spin up does not cancel with  
that of spin down in the presence of Zeeman splitting, and 
the total phase is equal to
$\int  {\bf p}_{up} \cdot d{\bf r} +\int {\bf p}_{down} \cdot d{\bf r}
=\int dr {\mu_B H}/{v_F}$.
Here $\mu_B$ is Bohr magnetor and $v_F$ is the Fermi velocity.
This leads to the sign oscillations of the Josephson coupling.
We assume the electrons in the metal are fully polarized but
neglect the spin polarization effect inside the grain
because the Zeeman splitting energy
scale is much smaller than the energy gap inside the 
grain. 
More specifically,
Josephson coupling between these two superconducting grains
is proportional to the cooper pair correlation function$^{[10, 11]}$ 
$ T \sum_n\{ \sigma_{y}^{\alpha\beta}
G^{\beta\gamma}_{\epsilon_n}({\bf r}_1, {\bf r}_2)
\sigma_y^{\gamma\delta}
G^{\delta\alpha}_{-\epsilon_n}({\bf r}_1, {\bf r}_2) \}$,
where $\sigma_y$ is the $y$ component 
Pauli matrix, $G^{\beta\gamma}_{\epsilon_n}$ is the Green function
in Mastubara representation with spin index $\beta, \gamma$.  
In the clean limit when $l \gg \mu_B H/v_F$ at $T=0$, the cooper
pair correlation function is  

\begin{equation}
\int d\epsilon d\theta\frac{d^2{\bf Q}}{(2\pi)^2}  
\frac{\cos\left[{\bf Q} \cdot ({\bf r}_1 -{\bf r}_2)\right]}
{2\epsilon + 2\mu_B H +{\bf v}_F \cdot {\bf Q}}
\tanh(\frac{\epsilon}{2kT}) 
\propto 
\cos\left(|{\bf r}_1 -{\bf r}_2|
\frac{2 \mu_B H}{v_F} -\frac{\pi}{4}\right)
\frac{1}{|{\bf r}_1 - {\bf r}_2|^2}
\end{equation} 
for $2D$ case at large distances. 
$\theta$ is the angle
between Fermi velocity and ${\bf Q}$.
The sign of the coupling oscillates with 
a period $v_F/\mu_B H$, which is much longer than the 
Fermi wave length, with which the sign of RKKY interaction
oscillates.

In disordered metals, 
for electron pairs to travel between grains
separated with a distance $L$, 
they must typically move along paths of length of order of $L^2/l$.
When $L^2/l \ll v_F/\mu_B H$, the sign is unpredictable.  
Indeed in the dirty limit when $L \gg \sqrt{D/\mu_B H} \gg l$,
the Josephson coupling averaged over the impurity configuration
is exponentially small 
$\exp(-{\sqrt{2}|{\bf r}_1 -{\bf r}_2|}/\sqrt{D/\mu_B H})$
while the typical amplitude of the coupling decays in the 
same way as in the pure limit,i.e. 
$\left\langle |E_J|\right\rangle \propto {|{\bf r}_1-{\bf r}_2|^{-2}}$. 
Therefore when the magnetic field increases, only the position
of the maximun of the distribution
function moves towards zero while the 
width of the distribution function barely changes.
This results in superconducting glass states.
Note that in principal the charging effect inside the grain will also lead
to superconducting glass phase as suggested in a recent experiment$^{[12]}$.
However in the metallic limit when the tunneling conductance
between the grain and the normal metal is much larger than
$e^2/\hbar$ such an effect is negligible. 

It is worth pointing out the mesoscopic pairing states
we discussed in this
paper are not originating from the inhomogeneity of the 
impurity concentrations. In fact, the 
range of the impurity potential is of atomic scale due to the
perfect screening in the metal and the fluctuation
of the impurity concentration within the area of coherence
length is inversely proportional to $\sqrt{n_{im} d\xi_0^2}$,
and becomes
negligibly small in the disordered metal. Here $n_{im}$ is the impurity
concentration. Most importantly, 
the long range coupling mentioned above, which leads to 
the superconducting glass state, is
not related to   
fluctuations of any local quantities. 
We are very grateful to 
B. Altshuler, L. I. Glazman, D. Huse, I. E. Smolyarenko,
B. Spivak for useful discussions.
F .Zhou is supported by  Princeton University.
C. Biagini is supported under PRA97-QTMD in Italy.
We also like to thank NEC Research Institute for its
hospitality.

\end{document}